\documentclass[12pt]{article}
\usepackage{eurosym,cite}
\usepackage[left=0.7in, right=0.7in, top=0.7in, bottom=0.7in]{geometry}
\usepackage{hyperref}
\usepackage{mathtools}
\usepackage{graphicx}
\usepackage{mathrsfs}
\usepackage[utf8]{inputenc}
\usepackage[T1]{fontenc}
\usepackage{mathrsfs}

\usepackage{tikz}
\usepackage{tikz-cd}
\usepackage{pgfplots}
\usepackage{color}

\begin{document}
	
	\begin{center}
		\textbf{\LARGE Entanglement-Driven Energy Exchange in a Two-Qubit Quantum Battery \\}
		\bigskip

 Ahmed A. Zahia$^{a}$ \textit{\footnote{corresponding author:ahmed.zahia@fsc.bu.edu.eg}}, 	M.Y.Abd-Rabbou$^{b,c}$ \textit{\footnote{e-mail:m.elmalky@azhar.edu.eg}}, Ahmed M. Megahed $^{a}$ \textit{\footnote{e-mail:ahmed.abdelbaqk@fsc.bu.edu.eg}}
	
	\small{ $^{a}${\footnotesize Department of Mathematics, Faculty of Science, Benha University, Benha, Egypt}\\
	$^{b}${School of Physics, University of Chinese Academy of Science, Yuquan Road 19A, Beijing, 100049, China}\\
	$^{c}${\footnotesize Department, Faculty of Science,
		Al-Azhar University, Nasr City 11884, Cairo, Egypt.}\\}

	\end{center}
	
	\begin{abstract}
		This study investigates the dynamics of quantum batteries (QBs), focusing on the pivotal role of quantum entanglement in mediating inter-cellular energy transfer within a two-cell configuration (two-qubit), wherein one cell is directly coupled to the charging source. Employing the Lindblad master equation to model the system's evolution, the influence of coherent state amplitudes, detuning, inter-cellular coupling strength, and dissipation rates on stored energy, energy fluctuations, concurrence-quantified entanglement, and their parametric interrelations is scrutinized. Our results indicate a direct correlation between the degree of entanglement and energy transfer efficiency between the qubits. Specifically, the stronger the entanglement between primary cell, which is connected to the charger, and secondary cell, the more effectively energy is transferred. This demonstrates that enhanced entanglement significantly facilitates energy transfer between the two qubits.
		
	\end{abstract}
\section*{Introduction}
The exploration of quantum phenomena for technological applications is a rapidly evolving research frontier, with QBs emerging as a particularly intriguing area of study \cite{alicki2013entanglement,ferraro2018high,campaioli2018quantum,henao2018role}. These quantum mechanical systems function as efficient energy storage devices, capitalizing on genuine quantum effects like entanglement and squeezing to enhance the performance of classical protocols by accelerating underlying dynamics  \cite{deffner2017quantum,giovannetti2003role}. While the advantages of quantum correlations in the charging and discharging processes of QBs have been discussed in theoretical contexts \cite{hovhannisyan2013entanglement,binder2015quantacell,campaioli2017enhancing}, recent
efforts have shifted towards concrete models that can be practically implemented in laboratories. Research has predominantly focused on maximizing stored energy, minimizing charging times, and enhancing average charging power \cite{le2018spin,andolina2018charger}. However, a truly effective QB must not only store significant energy but also deliver it efficiently for practical use, reflecting its thermodynamic capability to perform work. This duality presents a fascinating challenge: while quantum correlations can accelerate the charging process, they may simultaneously impose limits on the amount of work that can be extracted \cite{oppenheim2002thermodynamical,manzano2018optimal}, illustrating a nuanced interplay between the positive and negative effects of entanglement in energy storage.

Quantum entanglement is a fundamental concept in quantum mechanics, where two or more particles are so deeply interconnected that the state of one instantaneously affects the other, regardless of distance \cite{ryszard2009quantum,zyczkowski2001dynamics,zahia2023bidirectional}. This phenomenon has significant implications for technologies such as quantum communication, computing, and cryptography \cite{cleve1997substituting,hughes2000quantum,yin2020entanglement}. In quantum batteries, entanglement plays a crucial role in enhancing energy transfer and storage efficiency. By exploiting the quantum correlations
between entangled qubits, quantum batteries can achieve faster charging times and greater energy retention compared to classical systems \cite{kamin2020entanglement}. Studies have shown that entanglement allows for coherent energy transmission between qubits, optimizing energy flow \cite{shi2022entanglement}. Notably, entanglement enables collective quantum speed-up, making the charging process in quantum batteries exponentially faster than in non-entangled systems \cite{gyhm2024beneficial}. This quantum advantage, however, introduces challenges, as stronger entanglement can also lead to system instabilities, especially under the influence of dissipation or environmental noise \cite{carrega2020dissipative,xu2021enhancing}. As a result, balancing the strength of
quantum entanglement with system stability is essential for developing practical and efficient quantum batteries that can outperform classical energy storage technologies \cite{imai2023work}.

This paper investigates the effect of entanglement between two cells (qubits), emphasizing how this quantum correlation can facilitate energy transfer from one cell to another, effectively charging the second qubit. By delving into the mechanisms of entanglement, we aim to uncover the ultimate potential of quantum batteries and their applications in future quantum technologies. Utilizing the Lindblad master equation, we examine the impact of physical parameters—such as coherent state amplitudes, detuning parameter, coupling strength between the qubits, and dissipation rates—on entanglement, employing concurrence as a measurement tool. We assess energy charging and analyze energy fluctuations to evaluate stability within the context of these quantum systems.

The structure of this paper begins with Section \ref{section 1}, which introduces the Hamiltonian of the quantum battery model, detailing the interactions between cells in both $X$ and $Y$ directions, governed by the anisotropy parameter, and the coupling between the cavity field and primary cell. It also addresses the system’s dissipative dynamics using the Lindblad master equation. Section \ref{section 2}, the paper explores the energy dynamics of the quantum battery, quantifying stored energy and examining energy fluctuations over time, with a focus on system stability and how fluctuations affect energy transfer efficiency. Section \ref{section 3} investigates entanglement in energy transfer between qubits, specifically how entanglement—measured by concurrence—facilitates efficient energy transmission between the two-cell. A parametric analysis reveals how system parameters like coupling strength and detuning impact both energy and entanglement. Finally, Section \ref{section 4} concludes by emphasizing the balance between maximizing entanglement and ensuring system stability, and underscores the importance of tuning key parameters to optimize the performance of quantum batteries for future applications.

\section{Structure of the Quantum Battery System}\label{section 1}
\begin{figure}[h!]
	\centering
	\includegraphics[width=0.8\textwidth]{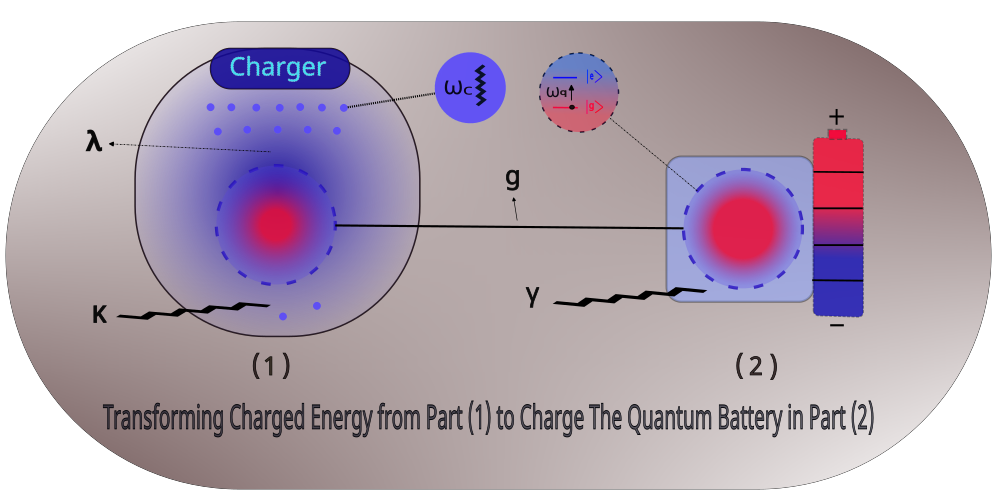}
	\caption{schematic diagram of a quantum battery system illustrating energy transfer between primary and secondary cells. Primary cell is coupled to the charger through the parameter $\lambda$, while primary and secondary cells are connected via the coupling constant $g$. The dissipation rates, $\kappa$ for the charger and $\gamma$ for the qubits, account for energy loss in the system. The charger operates at a frequency $\omega_c$, while the qubits oscillate at frequency $\omega_q$.} 
	\label{model}
\end{figure}
Consider a quantum model, depicted in Fig.~\ref{model}, which elucidates the transfer of energy from a charging source to a primary cell (a two-level quantum system), and its subsequent transmission to a secondary cell. We posit a quantum battery model comprising two cells: the primary cell is coupled to a radiation cavity field with a coupling strength $\lambda$, while the two cells interact via a coupling parameter $g$ along the $X$ and $Y$ axes. The directionality of this inter-cellular interaction is governed by the anisotropy parameter $\zeta$ \cite{breuer2002theory,lidar2003decoherence}. The Hamiltonian is defined given by ($\hbar=1$)
\begin{equation}
	\hat{H} = \omega_c \hat{a}^\dagger \hat{a} + \frac{\omega_q}{2} \sum_{j=1}^2 \hat{\sigma}_z^{(j)} 
	+ \lambda \left[\hat{\sigma}_+^{(1)} \hat{a} + \hat{\sigma}_-^{(1)} \hat{a}^\dagger \right] 
	+ \frac{g}{2} \left[ (1+\zeta) \hat{\sigma}_x^{(1)} \hat{\sigma}_x^{(2)} 
	+ (1-\zeta) \hat{\sigma}_y^{(1)} \hat{\sigma}_y^{(2)} \right].
\end{equation}
The operators $\hat{a}$ and $\hat{a}^\dagger$ represent the annihilation and creation operators of the cavity mode with frequency $\omega_c$. The qubits are described using Pauli matrices $\{\hat{\sigma}_i^{(j)}\}_{i=x,y,z}^{(j=1,2)}$ with frequency $\omega_q$. The anisotropy parameter $\zeta$ lies within $[-1, 1]$. When $\zeta = 0$, the interaction is isotropic, with equal coupling in both the X- and Y-directions. As $\zeta$ approaches $\pm 1$, the system becomes more anisotropic, with the interaction dominated by either the X- or Y-direction. The analysis focuses on $\zeta \in [0, 1]$ due to symmetry, where the anisotropy impacts the qubit interactions.

Though our investigation of this model, we scrutinize the system's dynamic interplay with its surroundings while adhering to the principle of energy conservation. To account for energy dissipation, we incorporate Lindblad dissipators into both the cavity and atomic subsystems to solve the system. The primary channels of energy loss within the system are the spontaneous emission rate, $\kappa$, characterizing the decay of the cavity field, and the atomic decay rate, $\gamma$. The system’s evolution is described by the Lindblad master equation, which is defined by \cite{zhang2016photon,nathan2020universal}
\begin{equation}
	\frac{d}{dt} \hat{\rho}(t) = -i[\hat{H}, \hat{\rho}(t)] 
	+ \kappa \mathcal{D}[\hat{a}] 
	+ \gamma \sum_{j=1}^2 \mathcal{D}[\hat{\sigma}_-^{j}],
\end{equation}
where $\mathcal{D}[\hat{O}] = \hat{O} \hat{\rho} \hat{O}^\dagger - \frac{1}{2} \{\hat{O}^\dagger \hat{O}, \hat{\rho}\}$ is the dissipator operator in terms of subsystem $\hat{O} $. For our system the Lindblad master equation generates the following system of ordinary differential equations
\begin{align}\label{3-a}
	\dot{\rho}_{11} &= -2\gamma \rho_{11} + i x_0 \rho_{12} + i x_0 \rho_{13} - i x_0 \rho_{21} + r_0^2 \kappa \rho_{22} + r_0^2 \kappa \rho_{33} - i x_0 \rho_{31},\notag \\
	\dot{\rho}_{12} &= i x_0 \rho_{11} + \Delta_0 \rho_{12} + i x_1 \rho_{14} + i x_0 \rho_{22} + r_0 r_1 \kappa \rho_{24} - i x_0 \rho_{32}=\rho_{21}^*,\notag \\
	\dot{\rho}_{13} &= i x_0 \rho_{11} + i x_1 \rho_{14} + \Delta_0 \rho_{13} + i x_0 \rho_{23} + r_0 r_1 \kappa \rho_{34} - i x_0 \rho_{31}=\rho_{31}^*,\notag \\
	\dot{\rho}_{14} &= i x_1 \rho_{12} + i x_1 \rho_{13} + \Delta_1 \rho_{14} - i x_0 \rho_{34} - i x_0 \rho_{24}=\rho_{41}^*,\notag \\
	\dot{\rho}_{22} &= \gamma \rho_{11} - i x_0 \rho_{12} + i x_0 \rho_{21} + \Delta_2 \rho_{22} + i x_1 \rho_{24} + \Delta_2^* \rho_{23} + r_1^2 \kappa \rho_{44},\notag \\
	\dot{\rho}_{23} &= -i x_0 \rho_{13} + i x_0 \rho_{21} + \Delta_2 \rho_{23} + \Delta_2^* \rho_{33} + i x_1 \rho_{43}=\rho_{32}^*,\notag \\
	\dot{\rho}_{24} &= \gamma \rho_{13} - i x_0 \rho_{14} + i x_1 \rho_{23} + \Delta_2^* \rho_{33} - i x_1 \rho_{43}=\rho_{42}^*,\tag{2-a} \\
	\dot{\rho}_{33} &= \gamma \rho_{11} - i x_0 \rho_{13} + \Delta_3 \rho_{33} + i x_1 \rho_{43} + \Delta_2^* \rho_{23} + r_1^2 \kappa \rho_{44},\notag \\
	\dot{\rho}_{34} &= \gamma \rho_{12} + i x_1 \rho_{14} + \Delta_3 \rho_{34} - i x_1 \rho_{43}=\rho_{43}^*,\notag \\
	\dot{\rho}_{44} &= \gamma \rho_{22} - i x_1 \rho_{24} + \gamma \rho_{33} - 2 r_1^2 \kappa \rho_{44},\notag 
\end{align}
where
\begin{align}
	x_0 &= \lambda \sqrt{n-1}, \quad x_1 = \lambda \sqrt{n+1}, \quad r_0 = \sqrt{n-1}, \quad r_1 = \sqrt{n},\notag \\
	\Delta_0 &= i \theta_1 - i \theta_0 - \frac{3\gamma + r_0^2 \kappa}{2}, \quad 
	\Delta_1 = i \theta_2 - i \theta_0 - (\gamma + r_1^2 \kappa),\notag \\
	\Delta_2 &= i g \zeta - \frac{r_0^2 \kappa}{2}, \quad \Delta_3 = -\gamma - r_0^2 \kappa,\notag \\
	\theta_0 &= (n-2) \omega_c + \omega_q, \quad \theta_1 = (n-1) \omega_c, \quad \theta_2 = n \omega_c + \omega_q.\notag
\end{align}

To solve the master equation, we assumed the initial state of the system is prepared by
\begin{equation}
	\hat{\rho}(0) = |gg, \alpha\rangle \langle gg, \alpha|, \quad 
	|\alpha\rangle = \sum_{n=0}^\infty c_n |n\rangle, \quad 
	c_n = e^{-\frac{|\alpha|^2}{2}} \frac{\alpha^n}{\sqrt{n!}},
\end{equation}
where the two cells are initially in the ground states $|gg\rangle$, while the charger in the coherent state, $|\alpha\rangle$, with a photon number $|\alpha|^2$, can be expressed as a superposition of Fock states $|n\rangle$.

Now, the solution of the system can be obtained by
\begin{equation}
	\hat{\rho}(t)=e^{tA}\hat{\rho}(0),
\end{equation}
since $A$ is the coefficients matrix of the entire system in (\ref{3-a}).

\section{Quantum Battery Energy Dynamics and Fluctuations}\label{section 2}
In this section, we delve into the dynamics of energy storage within QBs and scrutinize the inherent quantum fluctuations that characterize these systems. The discrete nature of energy levels in quantum systems results in a variability of stored energy within QBs, which can significantly influence their efficiency and reliability. A comprehensive understanding of these fluctuations, coupled with an analysis of energy evolution over time, is imperative for optimizing QB performance across diverse applications. This section explores the underlying mechanisms governing energy storage and fluctuations, while introducing analytical tools such as correlators to assess the stability and temporal behavior of the QB during the charging process.

\subsection{Stored Energy}
The quantitative assessment of energy stored within QBs is paramount for evaluating their operational efficacy. This energy metric, derived from collective excitation states within Hamiltonian frameworks, reflects the QB's capacity to perform work. The stored energy is influenced by a multitude of factors, including initial state preparation, charging protocols, and environmental interactions.  Enhancing energy management in QBs is particularly significant for applications such as quantum computing \cite{ho2018promise} and quantum sensing \cite{hatano2024wide}, where efficient energy utilization is paramount. The mathematical expression for the energy stored in the QB at a given time $t$ is expressed by \cite{crescente2020ultrafast}

\begin{equation}
	E(t) = \text{Tr}(\hat{\rho}(t)\hat{H}_{QB}) - \text{Tr}(\hat{\rho}(0)\hat{H}_{QB}),
\end{equation}
where $\hat{H}_{QB}=\frac{\omega_q}{2}\hat{\sigma}_z^{(2)}$.

\begin{figure}[!ht]
	\begin{center}
		\input{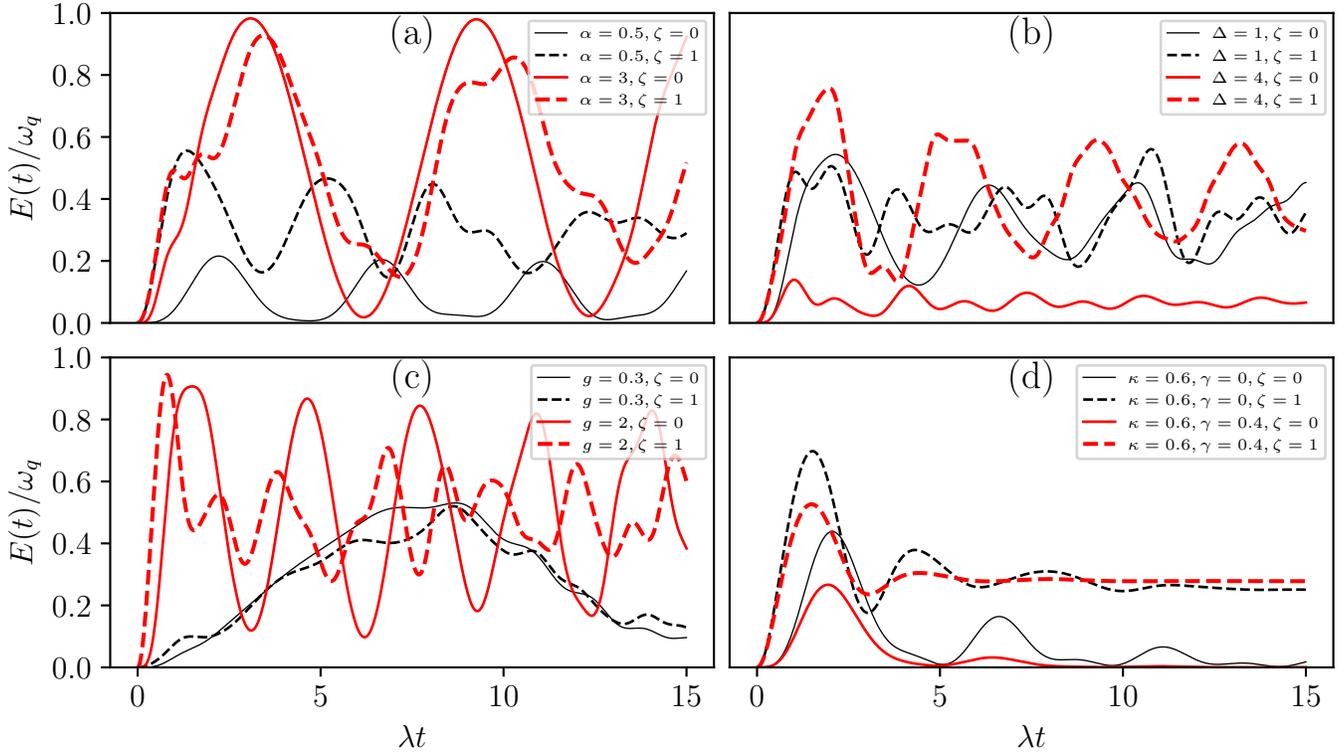} 
	\end{center}
	\vspace{-1 cm}
	\caption{The normalized stored energy $E(t)/ \omega_q$ as a function of the scaled time $\lambda t$ across isotropic interaction (solid-curve) and anisotropic interaction (dashed-curve), with system parameters: 
		(a) $\alpha$ is varied, while $\Delta = 0$, $g = 1$, $\kappa = \gamma = 0$.
		(b) $\Delta$ is varied, while $g = 1$, $\alpha = 2$, $\kappa = \gamma = 0$. 
		(c) $g$ is varied, while $\Delta = 0$, $\alpha = 2$, and $\kappa = \gamma = 0$. 
		(d) $\kappa$, and $\gamma$ are altered, while $\Delta = 0$, $\alpha = 1$, and $g = 1$.}
	\label{fig_energy}
\end{figure}

The temporal evolution of normalized stored energy $E(t)/ \omega_q$ within the non-interacting cell (second qubit) is depicted in Fig. \ref{fig_energy}. A detailed examination reveals how critical parameters—such as coupling strength ($g$), detuning ($\Delta$), dissipation rates ($\kappa$ and $\gamma$), and the anisotropy parameter ($\zeta$)—impact the system's charging and relaxation dynamics. In the resonance case, $\Delta=0$, and the absence of environment, Fig. \ref{fig_energy}(a) displays the energy dynamics under varying the coherent state amplitude $\alpha$ and anisotropy parameter $\zeta$. isotropic interaction $\zeta = 0$, the system exhibits moderate energy oscillations. Conversely, anisotropic interaction, $\zeta = 1$, enhances the energy exchange, as evidenced by sharper and higher energy peaks, indicating a more efficient charging process.  The first peak in the energy curves signify the initial charging period, while subsequent peaks correspond to the relaxation of stored energy. Higher values of $\alpha$ amplify the energy oscillations, demonstrating that larger coherent states facilitate faster and more robust energy transfer, particularly within the anisotropic interaction regime. Figure \ref{fig_energy}(b) depicts the impact of off-resonance conditions, with $\Delta = 1 \ \text{or} \ 4$, on the energy dynamics of an isolated system,  $\kappa = \gamma = 0$, and $\alpha=2$.  It is evident that as the detuning parameter, $\Delta$, increases,  the energy oscillations dampen, leading to a reduction in stored energy over time for the isotropic case. Conversely, the stored energy increases for the anisotropic case. This observation underscores the beneficial role of anisotropy in mitigating some of the detrimental effects of detuning. Compared to the results presented in Fig. \ref{fig_energy}(a), resonant conditions remain the optimal regime for energy storage. The impact of coupling strength $g$ and resonance condition is explored in Fig.\ref{fig_energy}(c). Stronger coupling ($g$ from 0.3 to 2) enhances energy exchange between qubits and the cavity field, increasing oscillation amplitude and energy transfer during the charging phase. With $\zeta = 1$, directional coupling further amplifies energy fluctuations, achieving sharper and more efficient transfer than the case of $\zeta = 0$. Figure \ref{fig_energy}(d) examines the influence of dissipation parameters $\kappa$ and $\gamma$ on energy retention. As anticipated, increasing dissipation results in a gradual attenuation of energy oscillations, with the stored energy decaying over time. This effect is most pronounced during the second peak, corresponding to the relaxation phase. Notably, when $\zeta = 1$, the directional interaction partially counteracts dissipation, enabling the system to retain more energy compared to the anisotropic case. This finding suggests that the isotropic configuration effectively mitigates energy loss by confining interactions to a specific direction, thereby reducing energy leakage during the relaxation phase.

From Fig. \ref{fig_energy}, a shift from isotropic interaction ($\zeta = 0$) to anisotropic interaction ($\zeta = 1$) enhances energy transfer and retention, particularly during the charging phase, by concentrating energy flow and improving efficiency. The resonance condition ($\Delta = 0$) maximizes stored energy, whereas detuning reduce energy transfer; however, anisotropy partially offsets this reduction. Strong coupling significantly accelerates the charging process and improves energy retention, with anisotropic interactions further optimizing overall performance. Although increased dissipation reduces stored energy, anisotropy mitigates this decay, thereby preserving system stability over time.

\subsection{Energy Fluctuation}
QBs exhibit energy storage fluctuations due to quantum discreteness. Understanding and controlling these fluctuations is crucial for reliable QB operation. Beyond average energy and optimal charging time \cite{campaioli2017enhancing}, temporal energy dynamics provide valuable insights. By analyzing energy correlators, researchers can track fluctuations and optimize QB performance during charging. Energy standard deviation is defined by \cite{crescente2020charging}

\begin{equation}
	\Sigma(t) = \sqrt{\text{Tr}\left(\delta(\hat{H}_{QB})^2 \hat{\rho}(0)\right) - \text{Tr}\left(\delta(\hat{H}_{QB}) \hat{\rho}(0)\right)^2},
\end{equation}
where $\delta(\hat{H}_{QB}) = \hat{H}(t)_{QB} - \hat{H}(0)_{QB}$, and $\hat{H}(t)_{QB} = \hat{U}^\dagger \hat{H}_{QB} \hat{U}$, the Hamiltonian evolved in time in the Heisenberg representation according to $\hat{H}$ is
\begin{equation}
	\hat{H}(t)_{QB} = e^{i\hat{H}t} \hat{H}_{QB} e^{-i\hat{H}t} = \hat{H}_{QB} + i t [\hat{H}_{QB}, \hat{H}] + \frac{(it)^2}{2!}[\hat{H}, [\hat{H}, \hat{H}_{QB}]] + \frac{(it)^3}{3!}[\hat{H}, [\hat{H}, [\hat{H}, \hat{H}_{QB}]]] + \cdots,
\end{equation}

\begin{figure}[!ht]
	\begin{center}
		\input{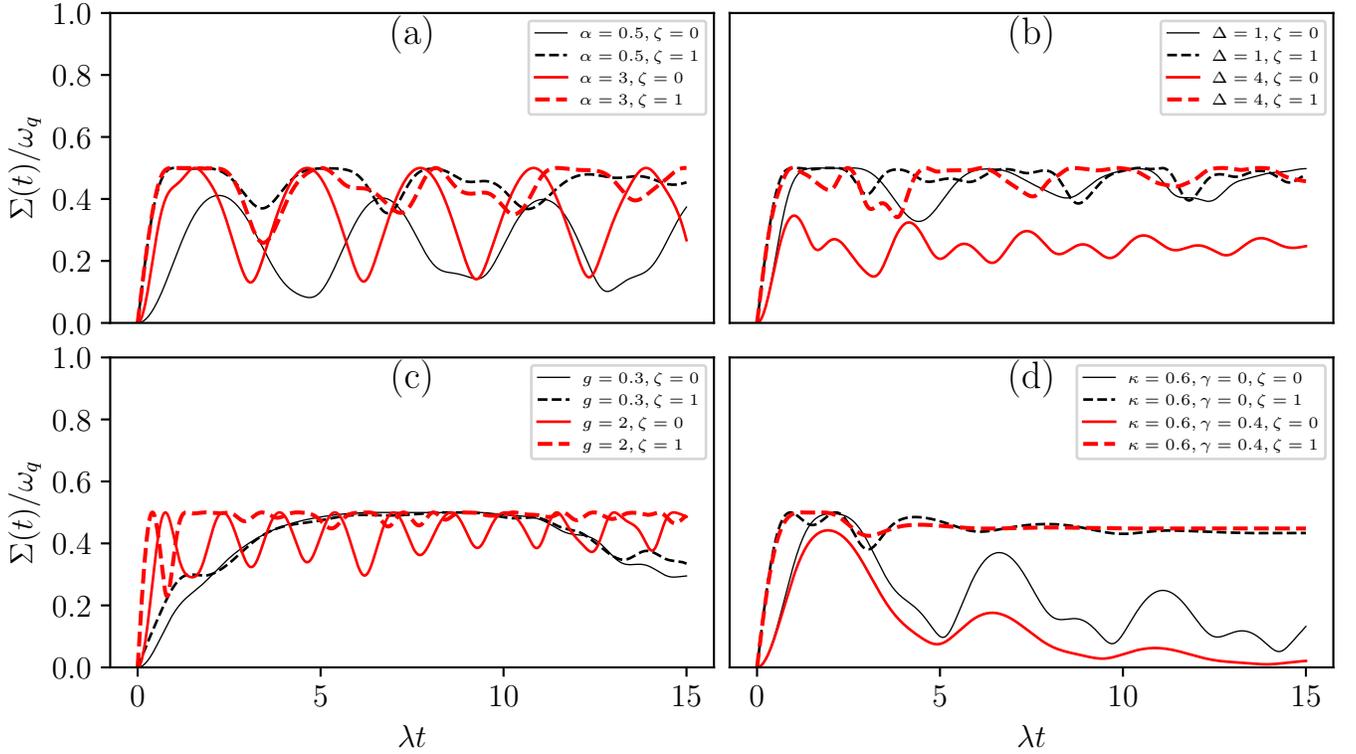} 
	\end{center}
	\vspace{-1 cm}
	\caption{The normalized energy fluctuation $\Sigma(t)/\omega_q$ with respect to scaled time $\lambda t$ with the same parameters that are displayed in Fig. \ref{fig_energy}.}
	\label{fig_fluctuation}
\end{figure}

The behavior of normalized energy fluctuations  $\Sigma(t)/2 \omega_q$ using the second qubit $\hat{H}_{QB}$ is shown in Fig. \ref{fig_fluctuation}. Energy fluctuations are a direct reflection of the inherent quantum mechanical uncertainty in the System, and controlling these fluctuations is crucial for achieving a stable and efficient quantum battery. Figure \ref{fig_fluctuation}(a) examines energy fluctuations for varying the coherent state intensity $\alpha$ and isotropy parameter $\zeta$. Lower $\alpha$ values result in smoother, less pronounced fluctuations, indicating enhanced stability. Conversely, higher $\alpha$ intensifies fluctuations, reflecting greater instability, particularly during the charging phase. Anisotropic interactions amplify these fluctuations compared to the isotropic case, suggesting that while larger coherent state amplitudes improve energy storage, they also increase variability, especially under anisotropic conditions. In Fig. \ref{fig_fluctuation}(b), the impact of detuning on energy fluctuations is displayed. With increased detuning ($\Delta = 1, 4$), the fluctuations become less structured and stability reduces over time, especially under anisotropic interaction ($\zeta = 1$), which exacerbates irregularity. These findings underscore the importance of resonance conditions in minimizing fluctuations and enhancing stability, as detuning compromises energy storage efficiency and increases instability. Fig. \ref{fig_fluctuation}(c) shows that weaker coupling ($g=0.3$) leads to smoother energy fluctuations, indicating greater stability. Stronger coupling ($g=2$) amplifies fluctuations, suggesting a trade-off between efficient energy transfer and stability. Anisotropy ($\zeta=1$) further exacerbates fluctuations, highlighting the potential for instability in systems with strong coupling and directional interactions. Figure \ref{fig_fluctuation}(d) examines the influence of dissipation parameters ($\kappa$, $\gamma$) on energy fluctuations. Increasing dissipation dampens fluctuations as energy is lost. Anisotropic interactions ($\zeta=1$) help maintain a more regular fluctuation pattern, suggesting potential stability benefits during relaxation.

The energy fluctuations $\Sigma(t)/\omega_q$ by our model reveals that larger coherent state amplitudes and strong coupling improve energy transfer efficiency but at the cost of heightened instability. Anisotropic interactions further exacerbate fluctuations by confining energy exchange, rendering the system more susceptible to instability. Ensuring resonance and minimizing dissipation are critical for maintaining stable energy storage, as off-resonance conditions and high dissipation significantly amplify instability. Achieving an optimal balance between energy transfer efficiency and system stability necessitates precise tuning of these parameters, which is vital for the effective operation of quantum batteries.

\section{Quantum Entanglement}\label{section 3}
In this section, we investigate quantum entanglement between two qubits using concurrence as the primary measure of entanglement \cite{gour2005family,wootters2001entanglement}.  For a given two-qubit density matrix $\hat{\rho}$, the concurrence $C(\hat{\rho})$ is defined as 
\begin{equation}
	C(\hat{\rho}) = \max(0, R_1 - R_2 - R_3 - R_4),
\end{equation}
where $R_1, R_2, R_3, R_4$ are the square roots of the eigenvalues, in decreasing order, of the matrix
\begin{equation}
	\tilde{\rho} = \hat{\rho} (\hat{\sigma}_y \otimes \hat{\sigma}_y) \hat{\rho}^* (\hat{\sigma}_y \otimes \hat{\sigma}_y).
\end{equation}
Here, $\hat{\rho}^*$ is the complex conjugate of $\hat{\rho}$, and $\hat{\sigma}_y$ is the Pauli-Y matrix. The concurrence ranges from 0 to 1, where 0 corresponds to a separable state and 1 corresponds to a maximally entangled state. 
\begin{figure}[!ht]
	\begin{center}
		\input{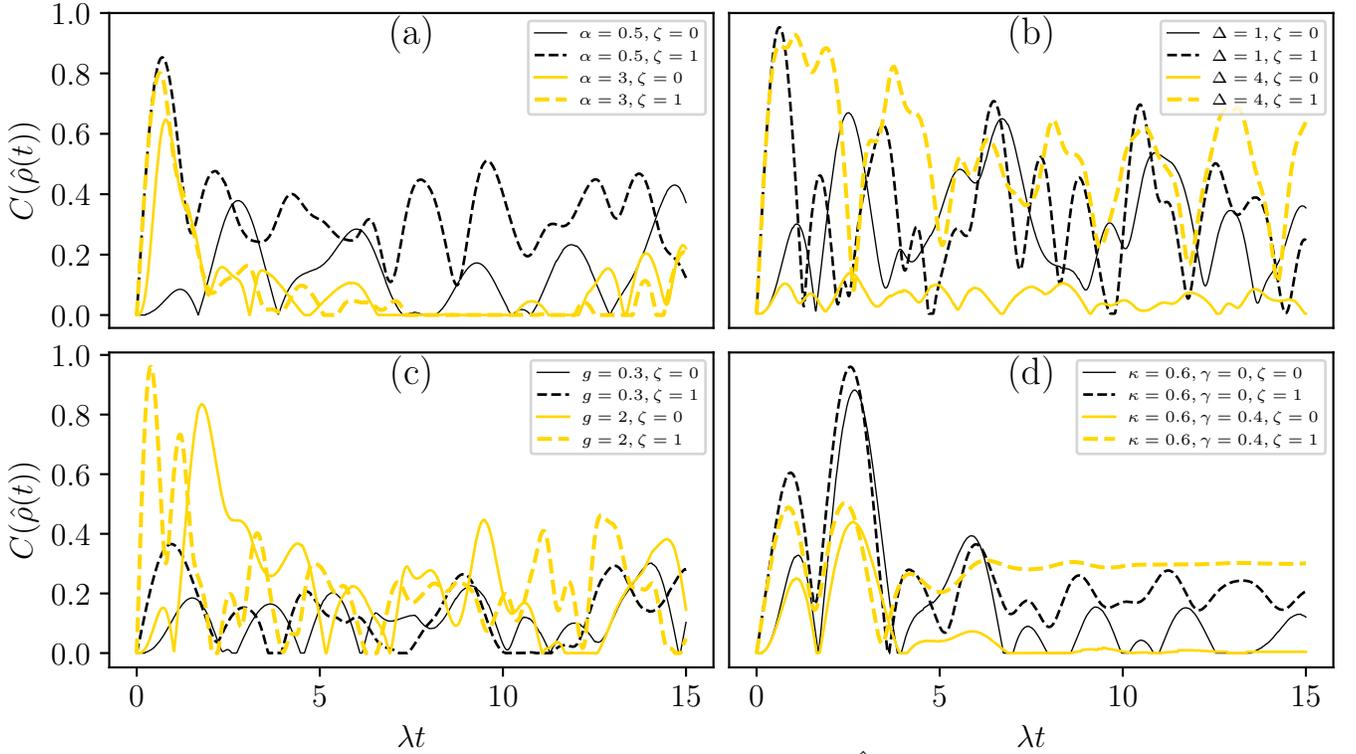} 
	\end{center}
	\vspace{-1 cm}
	\caption{The entanglement behaviour by the Concurrence $C(\hat{\rho(t)})$ as function of the scaled time $\lambda t$ with the same parameters that are displayed in Fig \ref{fig_energy}.}
	\label{fig_Concurence}
\end{figure}

The temporal entanglement between the two-cell is illustrated in Fig. \ref{fig_Concurence}, over the scaled time $\lambda t$ and various values of the system parameters.  In Fig. \ref{fig_Concurence}(a), we can see how varying the coherent state amplitude $\alpha$ and anisotropy $\zeta$ affects the entanglement. For lower $\alpha = 0.5$, the concurrence remains relatively low throughout the time evolution, indicating weaker partial entanglement. As $\alpha$ increases to 3, the entanglement increases significantly, especially in the early stages. However, the presence of anisotropy ($\zeta = 1$) introduces higher peaks in the concurrence values, indicating stronger but more unstable entanglement. Anisotropy amplifies the quantum correlations, enhancing the energy exchange between qubits but making the entanglement less consistent over time. Figure~\ref{fig_Concurence}(b) elucidates the influence of detuning parameter on the dynamical evolution of entanglement. Under off-resonant conditions, specifically when $\Delta = 1$, the concurrence exhibits a comparatively stochastic trajectory, fluctuating between partially entangled and separable states. Conversely, at an augmented detuning value of $\Delta = 4$, divergent entanglement dynamics manifest, contingent upon the system's anisotropy. Within the anisotropic regime, the system converges towards a partially entangled state. However, within the isotropic regime, a discernible attenuation of entanglement is observed, as well as generated separable states. In Fig.~\ref{fig_Concurence}(c), we observe the ramifications of coupling strength on concurrence. At a diminished coupling strength ($g = 0.3$), the concurrence remains attenuated, exhibiting intermittent phases of partial entanglement and disentanglement. Upon augmenting the coupling strength to $g = 2$, a discernible enhancement in inter-qubit entanglement ensues, evidenced by the amplified peaks in concurrence, albeit punctuated by brief interludes of disentanglement. The introduction of anisotropy ($\zeta = 1$) further accentuates these peaks, thereby suggesting that robust coupling, in conjunction with directional interactions, substantially elevates the degree of quantum entanglement, particularly during the nascent charging phase. Figure~\ref{fig_Concurence}(d) delineates the influence of dissipation on entanglement. In the absence of dissipation ($\kappa = 0$, $\gamma = 0$), the concurrence manifests relatively elevated and sustained maxima, indicative of robust quantum correlations between the qubits. Upon the introduction of dissipation ($\kappa = 0.6$, $\gamma = 0.4$), a precipitous decay in entanglement is observed, culminating in diminished peak amplitudes and a reduction in overall concurrence over time. Notably, when $\zeta = 1$, anisotropy serves to maintain comparatively elevated concurrence values, even amidst dissipative processes. This figure elucidates the precarious equilibrium between maximizing quantum entanglement and ensuring system stability. Augmented coherent state amplitudes and intensified coupling potentiate entanglement but concurrently exacerbate fluctuations and instability. Anisotropy, $\zeta = 1$, further amplifies entanglement, albeit at the expense of stability, particularly within systems exhibiting substantial detuning or dissipation. Preserving resonance, $\Delta = 0$, and minimizing dissipation are therefore paramount for sustaining robust, stable quantum correlations. Overall, subsequent to the initial peak, a decline in concurrence is observed, suggesting that while robust coupling expedites entanglement generation, it may concurrently engender decoherence over protracted durations.

\section{Energy Metrics and Quantum Entanglement}
This section interrogates the nexus between stored energy and its stability (fluctuations) as a function of quantum entanglement via parametric plots. Through systematic modulation of system parameters, the analysis visualizes the influence of quantum entanglement, quantified by concurrence, on both the energy storage capacity and stability of QBs. This investigation furnishes crucial insights into the interplay between entanglement and pivotal energy metrics, thereby underscoring the potential of entanglement optimization to enhance the performance and reliability of quantum systems.

\begin{figure}[!ht]
	\begin{center}
		\input{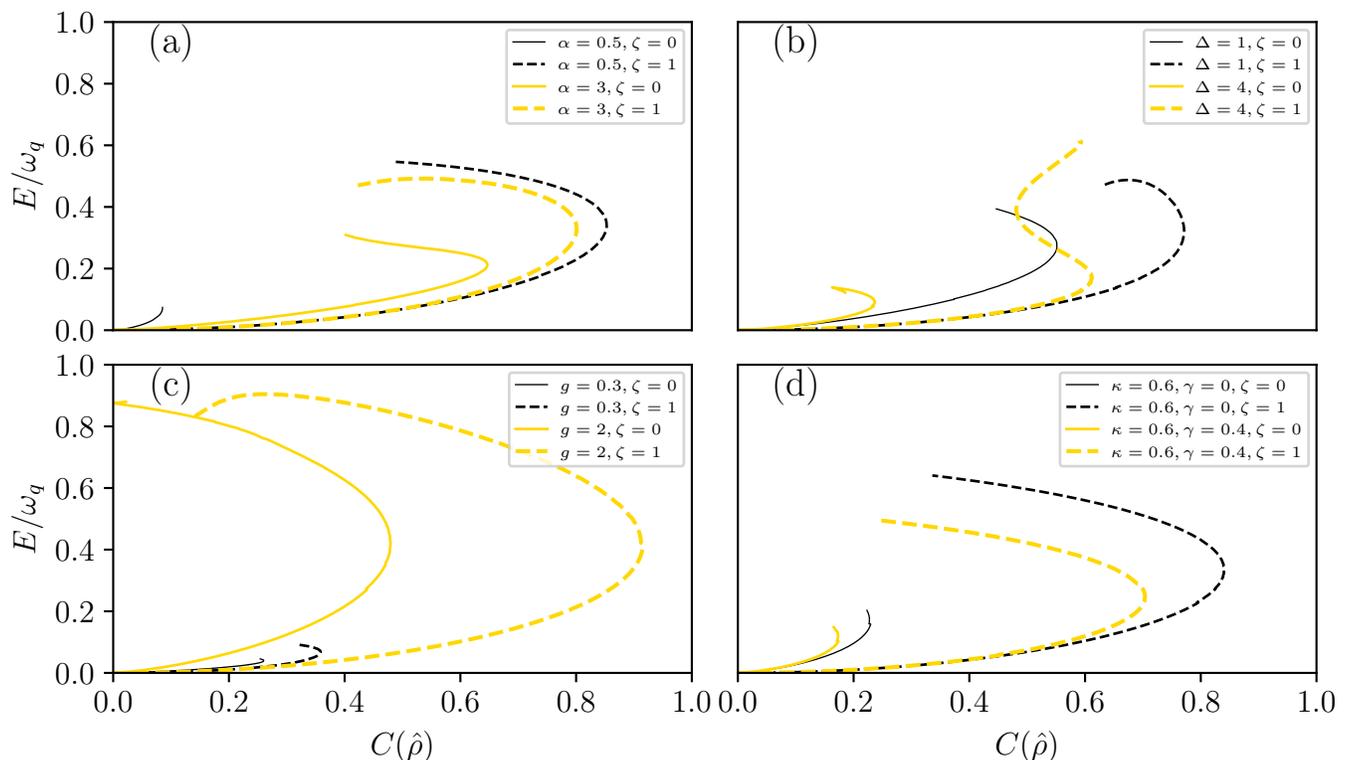} 
	\end{center}
	\vspace{-1 cm}
	\caption{The Parametric relation between the normalized stored energy $E(t)/\omega_q$ and the concurrence $C(\hat{\rho}(t))$ with respect to scaled time $\lambda t \in [0,1.3]$ (The first peak) with the same parameters that are displayed in Fig \ref{fig_energy}.}
	\label{fig_parametric_energy}
\end{figure}

Figure~\ref{fig_parametric_energy} illustrates the parametric relationship between the normalized stored energy $E(t)/\omega_q$ and the concurrence $C(\hat{\rho}(t))$, a measure of quantum entanglement, with respect to scaled time $\lambda t \in [0, 1.3]$, corresponding to the first peak in energy dynamics. The figure systematically examines the impact of various parameters on the interplay between energy storage and quantum entanglement. Figure~\ref{fig_parametric_energy}(a) investigates the role of coherent state amplitude $\alpha$ and anisotropy $\zeta$ on the energy-entanglement dynamics. At lower $\alpha$ values ($\alpha = 0.5$), the concurrence rises gradually with stored energy, eventually saturating at a plateau, which reflects diminishing entanglement gains at higher energy levels. Higher $\alpha$ values ($\alpha = 3$) significantly enhance both energy and entanglement, evidenced by a steeper rise and higher peaks in concurrence. However, this enhancement introduces fluctuations, particularly when anisotropy ($\zeta = 1$) is included, suggesting that directional interactions amplify entanglement at the expense of stability. This observation highlights the trade-off between enhanced performance and system robustness. Figure~\ref{fig_parametric_energy}(b) explores the influence of detuning $\Delta$ on the relationship between stored energy and concurrence. Under resonance conditions ($\Delta = 0$), the energy-entanglement correlation is smooth and reaches optimal levels. Conversely, under off-resonance conditions ($\Delta = 4$), both metrics diminish, with the anisotropic interaction ($\zeta = 1$) partially mitigating this reduction but at the cost of increased variability. This underscores the critical role of resonance in achieving stable and efficient energy transfer in quantum systems. Figure~\ref{fig_parametric_energy}(c) examines the effect of coupling strength $g$ on the energy-entanglement dynamics. For weak coupling ($g = 0.3$), the system demonstrates a stable but modest correlation between energy and concurrence. As $g$ increases ($g = 2$), both energy and entanglement are significantly enhanced, reflected in sharper peaks in concurrence. However, the anisotropic interaction ($\zeta = 1$) exacerbates fluctuations, leading to a rapid decline in entanglement beyond certain energy thresholds, revealing a potential instability in systems with strong coupling. Figure~\ref{fig_parametric_energy}(d) investigates the impact of dissipation parameters ($\kappa, \gamma$) on the system. In the absence of dissipation ($\kappa = 0, \gamma = 0$), the energy-entanglement correlation remains robust, with sustained maxima. However, introducing dissipation ($\kappa = 0.6, \gamma = 0.4$) significantly reduces both stored energy and concurrence. Anisotropic interactions ($\zeta = 1$) partially compensate for this loss but also amplify fluctuations, particularly at higher energy levels. These results emphasize the importance of minimizing dissipation to preserve entanglement and energy storage in practical implementations.

Overall, Figure~\ref{fig_parametric_energy} demonstrates the intricate balance required between key system parameters to optimize quantum battery performance. Enhanced coupling strength, larger coherent state amplitudes, and anisotropic interactions improve energy storage and entanglement but introduce challenges to system stability. Maintaining resonance and minimizing dissipation are essential for achieving a stable, positive correlation between energy and entanglement over extended durations. By carefully tuning these parameters, quantum batteries can achieve efficient and reliable energy transfer and storage.

\begin{figure}[!ht]
	\begin{center}
		\input{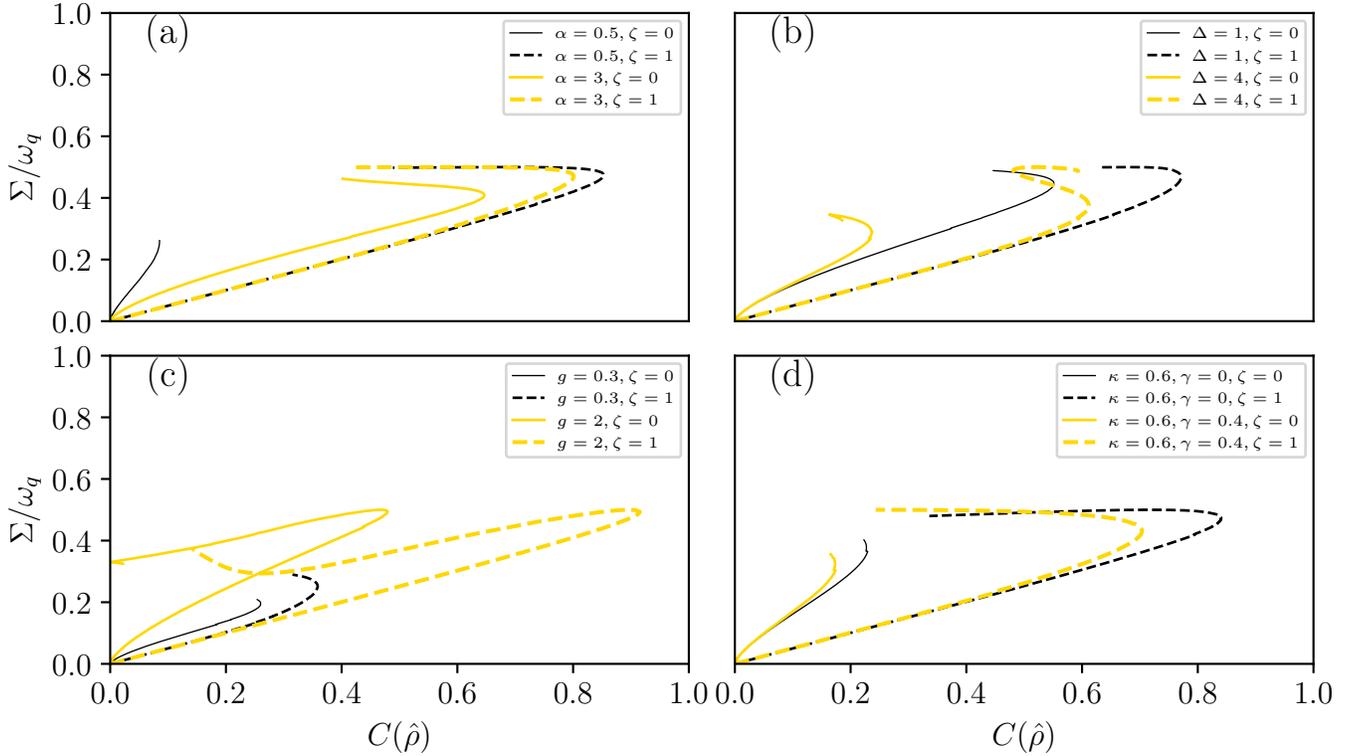} 
	\end{center}
	\vspace{-1 cm}
	\caption{The parametric relation between the normalized energy fluctuation $\Sigma(t)/\omega_q$ and the concurrence $C(\hat{\rho}(t))$ with respect to scaled time $\lambda t \in [0,1.3]$ (The first peak) with the same parameters that are displayed in Fig \ref{fig_energy}.}
	\label{fig_parametric_fluc}
\end{figure}
Figure~\ref{fig_parametric_fluc} delineates the parametric interrelation between normalized energy fluctuations, $\Sigma(t)/\omega_q$, and concurrence, $C(\hat{\rho}(t))$, which quantifies inter-qubit entanglement. These parametric plots are constructed for the scaled time corresponding to the first peak in energy dynamics, providing critical insights into the correlation between system stability, represented by energy fluctuations, and quantum entanglement under various conditions. Figure~\ref{fig_parametric_fluc}(a) investigates the relationship between energy fluctuations and concurrence under varying coherent state amplitude $\alpha$ and anisotropy $\zeta$. At lower $\alpha$ values ($\alpha = 0.5$), the concurrence exhibits a gradual increase with energy fluctuations, eventually reaching a plateau where further increases in $\Sigma(t)$ have negligible impact on entanglement. Conversely, at higher $\alpha$ values ($\alpha = 3$), the correlation becomes more pronounced, with a steeper rise in concurrence that peaks at higher fluctuation levels. The introduction of anisotropy ($\zeta = 1$) enhances this relationship, elevating concurrence values for a given level of energy fluctuations. However, this amplification comes at the expense of stability, as higher $\alpha$ values and anisotropy exacerbate fluctuations, reflecting a trade-off between improved energy transfer and reduced robustness. Figure~\ref{fig_parametric_fluc}(b) explores the influence of detuning $\Delta$ on the interrelation between energy fluctuations and concurrence. Under resonance conditions ($\Delta = 0$), the system exhibits a smooth and stable correlation, with concurrence increasing congruently with energy fluctuations. As detuning increases ($\Delta = 4$), this relationship becomes erratic, characterized by sharp peaks in concurrence at lower fluctuation levels followed by a subsequent decline. Anisotropy ($\zeta = 1$) mitigates some of the destabilizing effects of detuning by enhancing entanglement, though it also introduces additional variability at elevated fluctuation levels. These results emphasize the importance of maintaining resonance to sustain a stable energy fluctuation–entanglement relationship. Figure~\ref{fig_parametric_fluc}(c) illustrates the effect of coupling strength $g$ on the interrelation between energy fluctuations and concurrence. For weaker coupling ($g = 0.3$), the system demonstrates a modest yet stable correlation, with concurrence gradually increasing alongside energy fluctuations. As the coupling strength increases ($g = 2$), the concurrence rises more sharply, reflecting stronger inter-qubit interactions that enhance quantum entanglement. However, this enhancement introduces significant instability, as evidenced by a rapid decline in concurrence beyond a critical fluctuation threshold. The anisotropic interaction ($\zeta = 1$) further intensifies these effects, accelerating the initial growth in entanglement while amplifying instability at higher fluctuation levels. Figure~\ref{fig_parametric_fluc}(d) examines the impact of dissipation parameters ($\kappa, \gamma$) on the energy fluctuation–entanglement dynamics. In the absence of dissipation ($\kappa = 0, \gamma = 0$), the system maintains a robust and sustained correlation, with high concurrence persisting even at elevated fluctuation levels. However, as dissipation increases ($\kappa = 0.6, \gamma = 0.4$), this relationship weakens significantly. Both energy fluctuations and concurrence are reduced, with anisotropy ($\zeta = 1$) providing a partial recovery at lower fluctuation levels but amplifying variability and decay as dissipation intensifies. These findings underscore the detrimental effects of dissipation on system stability and entanglement, emphasizing the need for effective dissipation management in practical quantum battery implementations.

In summary, Figure~\ref{fig_parametric_fluc} highlights the delicate balance required between energy fluctuations and concurrence to optimize quantum battery performance. While enhanced coupling strength, larger coherent state amplitudes, and anisotropic interactions boost quantum entanglement, they also exacerbate energy fluctuations, leading to diminished stability. Resonance conditions are critical for maintaining a congruent and gradual ascent in both energy fluctuations and entanglement. In contrast, detuning and dissipation disrupt this balance, reducing the system’s ability to sustain robust quantum correlations. Carefully tuning system parameters—particularly anisotropy, coupling strength, and dissipation—remains essential for balancing stability and entanglement, ensuring effective and reliable energy storage and transfer to the secondary qubit.

\section{Conclusion}\label{section 4}
This study has investigated the pivotal role of quantum entanglement in mediating energy transfer between qubits, specifically in augmenting energy transmission from the primary cell, directly coupled to the charger, to the secondary cell, indirectly coupled to the charger but directly coupled to the primary cell, within a quantum battery system. The stored energy, energy fluctuations, entanglement, and their parametric interrelations have been scrutinized within this framework. The study demonstrated that key system parameters—including coherent state amplitude, detuning, coupling strength, anisotropy, and dissipation—played pivotal roles in modulating both entanglement strength and energy transmission efficiency.  
Quantum entanglement proved a crucial mechanism (in this system), not only facilitating but also optimizing inter-qubit energy flow. Enhanced entanglement correlated directly with more efficacious energy transfer, ensuring effective charging of secondary cell from the energy stored within primary cell.  Augmented coherent state amplitudes and intensified coupling enhanced entanglement, resulting in increased energy transfer, although this also engendered heightened fluctuations requiring careful management to maintain system stability. The introduction of anisotropy further potentiated entanglement and energy transmission but rendered the system more susceptible to instability. Resonance case proved essential for sustaining robust quantum correlations and optimizing energy transmission, whereas off-resonance conditions attenuated entanglement and reduced energy transfer effectiveness. Furthermore, dissipation parameter eroded entanglement, leading to diminished energy transmission between the two cells, underscoring the importance of minimizing dissipation to preserve both energy retention and quantum coherence. This investigation highlighted the precarious equilibrium between optimizing entanglement and maintaining stability within quantum batteries. 

\section*{Author contributions statement}
A .A. Z. prepared all figures and performed the mathematical calculations. M.Y. A.-R.  wrote the original draft. A.M. M. reviewed and edited the draft. All authors have read and agreed to the published version of the manuscript.    
\section*{Ethics declarations}
The authors declare no competing interests.

\section*{Availability of data and materials}
The used code of this study is available from the corresponding author upon reasonable
request.

\bibliographystyle{unsrt}
\bibliography{bm}

\begin{thebibliography}{10}

\bibitem{alicki2013entanglement}
Robert Alicki and Mark Fannes.
\newblock Entanglement boost for extractable work from ensembles of quantum
  batteries.
\newblock {\em Phys. Rev. E}, 87(4):042123, 2013.

\bibitem{ferraro2018high}
Dario Ferraro, Michele Campisi, Gian~Marcello Andolina, Vittorio Pellegrini,
  and Marco Polini.
\newblock High-power collective charging of a solid-state quantum battery.
\newblock {\em Phys. Rev. Lett.}, 120(11):117702, 2018.

\bibitem{campaioli2018quantum}
Francesco Campaioli, Felix~A Pollock, and Sai Vinjanampathy.
\newblock Quantum batteries.
\newblock {\em Thermodynamics in the Quantum Regime: Fundamental Aspects and
  New Directions}, pages 207--225, 2018.

\bibitem{henao2018role}
Ivan Henao and Roberto~M Serra.
\newblock Role of quantum coherence in the thermodynamics of energy transfer.
\newblock {\em Phys. Rev. E}, 97(6):062105, 2018.

\bibitem{deffner2017quantum}
Sebastian Deffner and Steve Campbell.
\newblock Quantum speed limits: from heisenberg’s uncertainty principle to
  optimal quantum control.
\newblock {\em J. Phys. A: Math. Theor.}, 50(45):453001, 2017.

\bibitem{giovannetti2003role}
Vittorio Giovannetti, Seth Lloyd, and Lorenzo Maccone.
\newblock The role of entanglement in dynamical evolution.
\newblock {\em Europhys. Lett.}, 62(5):615, 2003.

\bibitem{hovhannisyan2013entanglement}
Karen~V Hovhannisyan, Mart{\'\i} Perarnau-Llobet, Marcus Huber, and Antonio
  Ac{\'\i}n.
\newblock Entanglement generation is not necessary for optimal work extraction.
\newblock {\em Phys. Rev. Lett.}, 111(24):240401, 2013.

\bibitem{binder2015quantacell}
Felix~C Binder, Sai Vinjanampathy, Kavan Modi, and John Goold.
\newblock Quantacell: powerful charging of quantum batteries.
\newblock {\em New J. Phys.}, 17(7):075015, 2015.

\bibitem{campaioli2017enhancing}
Francesco Campaioli, Felix~A Pollock, Felix~C Binder, Lucas C{\'e}leri, John
  Goold, Sai Vinjanampathy, and Kavan Modi.
\newblock Enhancing the charging power of quantum batteries.
\newblock {\em Phys. Rev. Lett.}, 118(15):150601, 2017.

\bibitem{le2018spin}
Thao~P Le, Jesper Levinsen, Kavan Modi, Meera~M Parish, and Felix~A Pollock.
\newblock Spin-chain model of a many-body quantum battery.
\newblock {\em Phys. Rev. A}, 97(2):022106, 2018.

\bibitem{andolina2018charger}
Gian~Marcello Andolina, Donato Farina, Andrea Mari, Vittorio Pellegrini,
  Vittorio Giovannetti, and Marco Polini.
\newblock Charger-mediated energy transfer in exactly solvable models for
  quantum batteries.
\newblock {\em Phys. Rev. B}, 98(20):205423, 2018.

\bibitem{oppenheim2002thermodynamical}
Jonathan Oppenheim, Micha{\l} Horodecki, Pawe{\l} Horodecki, and Ryszard
  Horodecki.
\newblock Thermodynamical approach to quantifying quantum correlations.
\newblock {\em Phys. Rev. Lett.}, 89(18):180402, 2002.

\bibitem{manzano2018optimal}
Gonzalo Manzano, Francesco Plastina, and Roberta Zambrini.
\newblock Optimal work extraction and thermodynamics of quantum measurements
  and correlations.
\newblock {\em Phys. Rev. Lett.}, 121(12):120602, 2018.

\bibitem{ryszard2009quantum}
Horodecki Ryszard, Horodecki Pawel, Horodecki Michal, and Horodecki Karol.
\newblock Quantum entanglement.
\newblock {\em Rev. Mod. Phys}, 81(2):865--942, 2009.

\bibitem{zyczkowski2001dynamics}
Karol {\.Z}yczkowski, Pawe{\l} Horodecki, Micha{\l} Horodecki, and Ryszard
  Horodecki.
\newblock Dynamics of quantum entanglement.
\newblock {\em Phys. Rev. A}, 65(1):012101, 2001.

\bibitem{zahia2023bidirectional}
Ahmed~A Zahia, MY~Abd-Rabbou, Ahmed~M Megahed, and A-SF Obada.
\newblock Bidirectional field-steering and atomic steering induced by a magnon
  mode in a qubit-photon system.
\newblock {\em Sci. Rep.}, 13(1):14943, 2023.

\bibitem{cleve1997substituting}
Richard Cleve and Harry Buhrman.
\newblock Substituting quantum entanglement for communication.
\newblock {\em Phys. Rev. A}, 56(2):1201, 1997.

\bibitem{hughes2000quantum}
Richard~J Hughes and Colin~P. Williams.
\newblock Quantum computing: The final frontier?
\newblock {\em IEEE Intell. Syst. Their Appl}, 15(5):10--18, 2000.

\bibitem{yin2020entanglement}
Juan Yin, Yu-Huai Li, Sheng-Kai Liao, Meng Yang, Yuan Cao, Liang Zhang, Ji-Gang
  Ren, Wen-Qi Cai, Wei-Yue Liu, Shuang-Lin Li, et~al.
\newblock Entanglement-based secure quantum cryptography over 1,120 kilometres.
\newblock {\em Nature}, 582(7813):501--505, 2020.

\bibitem{kamin2020entanglement}
FH~Kamin, FT~Tabesh, S~Salimi, and Alan~C Santos.
\newblock Entanglement, coherence, and charging process of quantum batteries.
\newblock {\em Phys. Rev. E}, 102(5):052109, 2020.

\bibitem{shi2022entanglement}
Hai-Long Shi, Shu Ding, Qing-Kun Wan, Xiao-Hui Wang, and Wen-Li Yang.
\newblock Entanglement, coherence, and extractable work in quantum batteries.
\newblock {\em Phys. Rev. Lett.}, 129(13):130602, 2022.

\bibitem{gyhm2024beneficial}
Ju-Yeon Gyhm and Uwe~R Fischer.
\newblock Beneficial and detrimental entanglement for quantum battery charging.
\newblock {\em AVS Quantum Sci.}, 6(1), 2024.

\bibitem{carrega2020dissipative}
Matteo Carrega, Alba Crescente, Dario Ferraro, and Maura Sassetti.
\newblock Dissipative dynamics of an open quantum battery.
\newblock {\em New J. Phys.}, 22(8):083085, 2020.

\bibitem{xu2021enhancing}
Kai Xu, Han-Jie Zhu, Guo-Feng Zhang, and Wu-Ming Liu.
\newblock Enhancing the performance of an open quantum battery via environment
  engineering.
\newblock {\em Phys. Rev. E}, 104(6):064143, 2021.

\bibitem{imai2023work}
Satoya Imai, Otfried G{\"u}hne, and Stefan Nimmrichter.
\newblock Work fluctuations and entanglement in quantum batteries.
\newblock {\em Phys. Rev. A}, 107(2):022215, 2023.

\bibitem{breuer2002theory}
Heinz-Peter Breuer and Francesco Petruccione.
\newblock {\em The theory of open quantum systems}.
\newblock Oxford University Press, USA, 2002.

\bibitem{lidar2003decoherence}
Daniel~A Lidar and K~Birgitta~Whaley.
\newblock Decoherence-free subspaces and subsystems.
\newblock In {\em Irreversible quantum dynamics}, pages 83--120. Springer,
  2003.

\bibitem{zhang2016photon}
Yang Zhang, Jun Zhang, and Chang-shui Yu.
\newblock Photon statistics on the extreme entanglement.
\newblock {\em Sci. Rep.}, 6(1):24098, 2016.

\bibitem{nathan2020universal}
Frederik Nathan and Mark~S Rudner.
\newblock Universal lindblad equation for open quantum systems.
\newblock {\em Phys. Rev. B}, 102(11):115109, 2020.

\bibitem{ho2018promise}
Alan Ho, Jarrod McClean, and Shyue~Ping Ong.
\newblock The promise and challenges of quantum computing for energy storage.
\newblock {\em Joule}, 2(5):810--813, 2018.

\bibitem{hatano2024wide}
Yuji Hatano, Junya Tanigawa, Akimichi Nakazono, Takeharu Sekiguchi, Shinobu
  Onoda, Takeshi Ohshima, Takayuki Iwasaki, and Mutsuko Hatano.
\newblock A wide dynamic range diamond quantum sensor as an electric vehicle
  battery monitor.
\newblock {\em Philos. Trans. R. Soc., A}, 382(2265):20220312, 2024.

\bibitem{crescente2020ultrafast}
Alba Crescente, Matteo Carrega, Maura Sassetti, and Dario Ferraro.
\newblock Ultrafast charging in a two-photon dicke quantum battery.
\newblock {\em Phys. Rev. B}, 102(24):245407, 2020.

\bibitem{crescente2020charging}
A~Crescente, M~Carrega, M~Sassetti, and D~Ferraro.
\newblock Charging and energy fluctuations of a driven quantum battery.
\newblock {\em New J. Phys.}, 22(6):063057, 2020.

\bibitem{gour2005family}
Gilad Gour.
\newblock Family of concurrence monotones and its applications.
\newblock {\em Phys. Rev. A: At., Mol., Opt. Phys.}, 71(1):012318, 2005.

\bibitem{wootters2001entanglement}
William~K Wootters.
\newblock Entanglement of formation and concurrence.
\newblock {\em Quantum Inf. Comput.}, 1(1):27--44, 2001.

\end{thebibliography}

\end{document}